\documentclass[journal=jpcafh,manuscript=article]{achemso}

\usepackage[utf8]{inputenc}
\usepackage[OT1]{fontenc}

\usepackage{graphicx}
\graphicspath{{./}{./IMAGES/}}
\usepackage{amssymb}
\usepackage{amsmath}
\usepackage{amsfonts}

\usepackage{natbib}

\title{Connecting elasticity and effective interactions of neutral microgels: the validity of the Hertzian model}

\author{Lorenzo Rovigatti}
\affiliation{Dipartimento di Fisica, {\em Sapienza} Universit\`a di Roma, Piazzale A. Moro 2, 00185 Roma, Italy}
\alsoaffiliation{CNR-ISC, Uos Sapienza, Piazzale A. Moro 2, 00185 Roma, Italy}
\email{lorenzo.rovigatti@uniroma1.it}

\author{Nicoletta Gnan}
\affiliation{CNR-ISC, Uos Sapienza, Piazzale A. Moro 2, 00185 Roma, Italy}
\alsoaffiliation{Dipartimento di Fisica, {\em Sapienza} Universit\`a di Roma, Piazzale A. Moro 2, 00185 Roma, Italy}

\author{Andrea Ninarello}
\affiliation{CNR-ISC, Uos Sapienza, Piazzale A. Moro 2, 00185 Roma, Italy}
\alsoaffiliation{Dipartimento di Fisica, {\em Sapienza} Universit\`a di Roma, Piazzale A. Moro 2, 00185 Roma, Italy}

\author{Emanuela Zaccarelli}
\affiliation{CNR-ISC, Uos Sapienza, Piazzale A. Moro 2, 00185 Roma, Italy}
\alsoaffiliation{Dipartimento di Fisica, {\em Sapienza} Universit\`a di Roma, Piazzale A. Moro 2, 00185 Roma, Italy}
\email{emanuela.zaccarelli@cnr.it}

\usepackage{hyperref}
\hypersetup{
allcolors=black
}

\begin{document}

\begin{abstract}
An important open problem in materials science is whether a direct connection exists between single-particle elastic properties and the macroscopic bulk behavior. Here we address this question by focusing on the archetype of soft colloids: thermoresponsive microgels. These colloidal-sized polymer networks are often assumed to interact through a simple Hertzian potential, a classic model in linear elasticity theory. By developing an appropriate methodology, that can be generalized to any kind of soft particle, we are able to calculate all the elastic moduli of non-ionic microgels across their volume phase transition (VPT). Remarkably, we reproduce many features seen in experiments, including the appearance of a minimum  of the Poisson's ratio close to the VPT. By calculating the particle-particle effective interactions and the resulting collective behavior, we find that the Hertzian model works well up to moderate values of the packing fraction.
\end{abstract} 

\maketitle

\section{Introduction}

Colloidal suspensions have been used for decades as model systems for investigating fundamental condensed matter phenomena~\cite{pusey1986phase, aastuen1986nucleation, trappe2001jamming, pham2002multiple, mattsson2009soft, hunter2012physics}. Compared to atomic and molecular systems, colloids have much larger characteristic time- and length-scales, which makes them more accessible from an experimental point of view. In this context, the most iconic (and probably studied) soft matter system is certainly hard spheres~\cite{pusey1986phase,hansen2013theory}.
Being the prototype of athermal particles, hard spheres have a single control parameter that fully determines their behaviour, \textit{i.e.} the packing fraction. Despite the unique simplicity of the system, hard spheres exhibit a non-trivial behaviour that features crystallisation~\cite{aastuen1986nucleation, auer2001prediction,gasser2001real,sanz2014avalanches,thorneywork2017two} as well as glass and jamming transitions~\cite{van1998measurement,kegel2000direct, brambilla2009probing,parisi2010mean}.

By increasing the complexity of the system, and hence going beyond hard spheres, the range of possibilities is greatly enlarged. Indeed, in the soft-matter realm the mutual interactions between building blocks can be tuned to a high degree, making it possible to work with particles that effectively attract each other (and hence can, for instance, drive the appearance of a gas-liquid separation and of the associated critical point~\cite{And02aNature}) 
or that interact through non-spherical potentials~\cite{bianchi2006phase, wang2012colloids, ruzicka2011fresh} and can give rise to, for instance, exotic crystals or liquid-crystal phases~\cite{chen2011directed, wang2017colloidal, van2000liquid}. However, for the majority of these cases the shape and size of the particles do not depend on the external control parameters (temperature, packing fraction, external fields, \textit{etc.}) and can thus be effectively considered as undeformable objects. Adding internal degrees of freedom, and thereby making the particles \textit{soft} and \textit{responsive}, completely changes their behaviour~\cite{vlassopoulos2014tunable}. Far from being a mere academic curiosity, the role played by the softness in determining the overall behaviour of a system is fundamental in many diverse fields, ranging from biology~\cite{zaccai2000soft, shelby2003microfluidic} to fundamental physics~\cite{mattsson2009soft,PhysRevLett.114.098303,van2017fragility,gnan2018microscopic}.

Among the most used soft colloids, microgels have recently gained significant interest in condensed matter physics. Microgels are particles made of crosslinked polymer networks that can be responsive to external stimuli such as temperature or pH changes~\cite{fernandez2011microgel}. Indeed, as the external conditions change, the polymeric nature of the particle reacts by adjusting its internal structure. A well-known phenomenon associated to thermoresponsive microgels, usually based on Poly-N-isopropylacrylamide (PNIPAM), is the so-called volume phase transition (VPT), a sharp change of the particle size due to the variation of temperature, and hence of the quality of the solvent as felt by the polymers that constitute the network~\cite{FloryJcp11}. From an applicative standpoint, the responsiveness of microgels can be used to produce smart materials such as lenses or photonic crystals~\cite{debord2000thermoresponsive}, to stabilize emulsions~\cite{Liu384} or can be employed in biomedicine as drug-delivery carriers~\cite{oh2008development}, in food industry~\cite{shewan2013review}, in water purification~\cite{menne2014temperature} or in chemical sensing~\cite{prodi2005luminescent}.
From a fundamental perspective, the possibility of finely tuning the size of the particle makes it possible to control the packing fraction \textit{in situ}, without changing the number density of the samples. Such a high degree of control is very important to investigate phenomena that are very sensitive to packing, such as glass and jamming transitions~\cite{yunker2014physics,rossi2015shape,scotti2016role}. 
However, varying the size of particles, \textit{e.g.} through a temperature change, in turn affects the internal structure and the interactions between the particles. Furthermore, different synthetic routes lead to different microscopic morphologies, ranging from rather homogeneous particles\cite{pelton2011microgels} to heterogeneous core-corona architectures\cite{berndt2006temperature} and even to hollow microgels~\cite{scotti2018hollow}. 
Since the inner structure determines the single-particle properties, and hence the bulk behaviour of the system, a realistic description of the particles should take into account at least some aspects of the polymeric nature of the microgels in order to capture the way in which they interact with the environment and among themselves.

The complexity of a suitable theoretical description can be decreased by systematically reducing the particles' internal degrees of freedom~\cite{likos2001effective}. For microgels the most extreme coarse-graining approach is to model them as spheres interacting through effective pair interactions, completely neglecting their internal degrees of freedom. Even though this strategy is strictly valid only in the dilute regime, it has been shown to yield satisfactory results in a wider concentration region~\cite{mohanty2014effective}. In the regime where a pair-interaction description is supposed to work, it might also be appropriate to approximate the effective potential between two microgels with the classic Hertzian interaction, derived by computing the force acting between two elastic spheres in the framework of the classical elasticity theory (CET)~\cite{landau1986theory}. Indeed, results based on the inversion of radial distribution functions of dilute suspensions measured with confocal microscopy experiments have found functional forms that can be well represented with a Hertzian potential~\cite{zhang2009thermal}. Under the CET assumptions, the Hertzian effective potential depends only on the geometry and on the zero-stress elastic moduli of the materials that compose the objects~\cite{landau1986theory}, which thus incorporate the information about the particle's architecture, chemical composition and interaction with the solvent. Consequently, it should be possible to directly derive the effective interactions, valid at least at low concentrations, by measuring the single-particle elastic properties. These can be accessed by AFM~\cite{hashmi2009mechanical,C8NR02911C}, capillary micromechanics~\cite{voudouris2013micromechanics} or osmotic pressure~\cite{wyss_osmotic} experiments. These measurements should then be complemented by the evaluation of the effective interactions in order to validate the Hertzian theory. Such an experimental effort has not been carried out yet, probably due to the difficulty to perform both sets of experiments on the very same particles. However, a recent work~\cite{maxime_nat_comm} has investigated the effective interactions of microgels in suspension with small depletants, showing that, when an additional attraction is present, the standard Hertzian model is not valid even at moderate concentrations.

Turning to the numerical side, to date there are no complete studies on the elasticity of single-particle microgels, as this requires a realistic mesoscopic modelling of the underlying polymer network which, till very recently, was missing. Indeed, single microgels have been modelled in the past few years as coarse-grained polymer networks built by placing the crosslinkers on a regular lattice (\textit{e.g.} a four-coordinated diamond lattice) and joining them with same-length chains. This model has been used to assess the properties of both neutral and charged microgels~\cite{kobayashi2014structure,schroeder2015electrostatic,Ahualli2017,keidel2018time}
and more recently also to investigate the behaviour of microgels with more complicated architectures~\cite{scotti2018hollow,C8SM00170G}. In order to get rid of the crystalline structure that underlies diamond-generated networks, new procedures have been recently developed to generate disordered microgels with more realistic inner structures~\cite{kamerlin2016collapse,gnan2017silico,nikolov2018mesoscale,moreno2018computational,rovigatti2019numerical}. Interestingly, a recent simulation study reported non-Hertzian effective interaction potentials between neutral diamond-based microgels~\cite{Ahualli2017}, at odds with direct and indirect experimental evidence~\cite{mohanty2014effective,zhang2009thermal}. In this work, however, elastic moduli were not calculated but fixed with ad-hoc assumptions.

In the Hertzian approach microgels are treated as elastic spheres of diameter $\sigma$,  Young's modulus $Y$ (or bulk modulus $K$) and  Poisson's ratio $\nu$. However, the calculation of the elastic moduli of single microgels is a challenging task. Indeed, the crucial point is that, in order to verify the Hertzian picture, and thus to examine the link between single-particle elasticity and bulk behavior, the simultaneous knowledge of at least two elastic moduli is required.
Recent results have reported the dependence of the bulk modulus across the volume phase transition~\cite{nikolov2018mesoscale}, but  estimates of the other moduli were not provided and a connection with the architecture of the polymer network has not been established yet.
The main difficulty of setting such a crucial link is posed by the actual calculation of the elastic moduli at the level of a single (finite-sized) object. In contrast with biology and biophysics, where much effort has been devoted to the numerical investigation of the elastic properties of, for instance, cells~\cite{boal1994computer, boey1998simulations}, membranes~\cite{paulose2012fluctuating} and protein assemblies~\cite{aggarwal_pre,aggarwal_2018}, similar studies on artificial soft-matter particles have been scarce~\cite{riest2015elasticity}. Here we fill this gap by adapting the method developed by Aggarwal \textit{et al.} in the context of virus capsids~\cite{aggarwal_pre} and extending it to compute the elastic properties of soft colloids. 

To this aim we perform simulations of neutral microgels generated with a sophisticated method that we recently devised and that was shown to build realistic PNIPAM microgels~\cite{gnan2017silico}. Here the network is completely disordered and possesses the typical inhomogeneous core-corona structure, as experimentally measured by neutron scattering experiments. We calculate all elastic moduli of these microgels as a function of the crosslinker concentration and we also determine the effective interactions between two of these microgels. We are then able to directly connect the elastic properties to the effective potential, clearly demonstrating that, at small deformations, the Hertzian potential is valid. However, as the two microgels become closer and closer, a situation that occurs at high nominal packing fractions $\zeta = \pi\sigma^3 \rho/6$ of the suspension, where $\rho$ is the number density, the CET assumptions clearly break down and the effective interactions are found to significantly deviate from the Hertzian law. By performing additional simulations of coarse-grained microgels interacting with the predicted Hertzian potential and with the calculated effective potential, we finally assess that the range of validity of the Hertzian potential should hold up to packing fractions smaller than $\zeta \approx 1$. Since the majority of the interesting phenomena arising in microgel suspensions, such as jamming or dynamical arrest, happen at even higher values of the packing fraction, our results demonstrate that, in this regime, more complicated processes than just simple two-body elastic repulsions come into play. Therefore, different models and methodologies will have to be adopted to correctly describe microgels under very dense conditions.

\section{Materials and Methods}

The microgels that we use in our simulations are self-assembled disordered, fully-bonded networks that possess a well-defined core-corona structure~\cite{gnan2017silico}. The method we use to generate microgel conformations yields particles whose inner structure is fully compatible with that of PNIPAM microgels of radius $\approx 400$ nm, synthesised through precipitation polymerization, as revealed by comparison with small angle X-ray scattering measurements~\cite{ninarello2019advanced}. The numerical microgel configurations are composed of $\approx 5000$ beads and are generated by simulating the self-assembly of a binary mixture composed of divalent and tetravalent particles confined in a sphere of fixed radius $Z = 25$~$\sigma_m$~\cite{gnan2017silico}, where $\sigma_m$ is the monomer diameter. This specific value of $Z$ has been shown to yield microgels with internal structures compatible with experimental measurements~\cite{ninarello2019advanced}. Once the network has formed, we fix the topology of the resulting structure and we remove the confinement. We evaluate the influence of the crosslinker concentration $c$ by generating three different classes of microgels with $c = 3.2\%$, $5.0\%$ and $10\%$, respectively.

In order to evaluate the elastic properties of single microgels and the microgel-microgel interaction potential we run constant-temperature molecular dynamics simulations.
The microgel monomers interact with a classical bead-spring model for polymers\cite{grest1986molecular}, in which bonded monomers interact \textit{via} the sum of a Weeks-Chandler-Andersen (WCA), $V_{\rm WCA}(r)$, and a Finite-Extensible-Nonlinear-Elastic (FENE), $V_{\rm FENE}(r)$, potentials:
\begin{eqnarray*}\label{wca}
V_{\rm WCA}(r) &=&\begin{cases}
    4\epsilon\left[\left(\frac{\sigma_m}{r}\right)^{12}-\left(\frac{\sigma_m}{r}\right)^{6}\right]+\epsilon & \text{if $r \le 2^{\frac{1}{6}}\sigma_m$}\\
    0 & \text{otherwise}
  \end{cases}\\
V_{\rm FENE}(r) &=&-\epsilon k_FR_0^2\ln(1-(\frac{r}{R_0\sigma_m})^2)      \text{ if $r < R_0\sigma_m$}
\end{eqnarray*}
with $k_F=15$ a dimensionless spring constant and $R_0=1.5$ the maximum extension value of the bond. Non-bonded monomers only experience a repulsive WCA potential. In addition, 
 the thermoresponsivity of PNIPAM microgels is mimicked by adding an attractive term that implicitly controls the quality of the solvent\cite{soddemann2001generic,verso2015simulation}:
\begin{equation}\label{alpha}
V_{\alpha}(r)=\begin{cases}
    -\epsilon\alpha & \text{if $r \le 2^{\frac{1}{6}}\sigma_m$}\\
    \frac{1}{2}\alpha\epsilon\left[\cos\left(\delta\left(\frac{r}{\sigma_m}\right)^2+\beta\right)-1\right] & \text{if $2^{\frac{1}{6}}<r\le R_0\sigma_m$}\\
    0. & \text{otherwise}
  \end{cases}
\end{equation}
where $\delta=\pi(2.25-2^{1/3})^{-1}$ and $\beta=2\pi-2.25\delta$~\cite{soddemann2001generic}. 
With this potential the solvophobicity of the polymers is controlled by the parameter $\alpha$, which plays the role of the temperature: when $\alpha=0$, monomers do not experience any mutual attraction, as under good solvent conditions. As $\alpha$ increases the monomers become more and more attractive, mimicking bad solvent conditions. This potential was shown to be able to accurately reproduce the phenomenology of the volume phase transition\cite{gnan2017silico,rovigatti2017internal,moreno2018computational} which is found to occur, independently of the internal microgel topology, at $\alpha \sim 0.6$.

Reduced temperature $T^*=k_BT/\epsilon=1.0$ is enforced by using a modified Andersen thermostat~\cite{russo2009reversible}, where $k_B$ is the Boltzmann constant, $T$ the temperature and $\epsilon$ is the monomer interaction strength. The equations of motion are integrated with a velocity-Verlet algorithm using a reduced timestep $\Delta t^* =\Delta t \sqrt{\epsilon/(m\sigma_m)}= 0.002$ where $m$ is the mass of the monomer.
 Length, mass and energy are given in units of $\sigma_m$, $m$ and $\epsilon$, respectively. We run simulations %on single GPUs 
 for $10^8$--$10^9$ time steps, depending on the microgel investigated.

We calculate the microgel-microgel effective interactions using two distinct and complementary methods. We simulate two microgels at different center-to-center separation  $r$. To tackle the large-separation region we use a generalised Widom insertion scheme~\cite{mladek2011pair}, which is very efficient in sampling the small-deformation regime. At smaller separations (or, equivalently, at larger deformations) we use umbrella sampling by adding an harmonic biasing potential acting on the centres of mass of the two microgels~\cite{roux1995calculation}. The resulting effective potential is calculated as,
\begin{equation}
V(r) = -k_B T \ln g(r)
\label{eq:pot}
\end{equation}
\noindent
where $g(r)$ is the radial distribution function. 

In the Hertzian approximation, two elastic spheres interact with an effective potential that is induced by the elastic mechanical deformation as~\cite{landau1986theory}:
\begin{equation}
\label{eq:hertzian_fit}
V_H(r) = \frac{2 Y \sigma^3}{15 (1 - \nu^2)} \left( 1 - \frac{r}{\sigma}\right)^\frac{5}{2}\theta(1-r/\sigma),
\end{equation}
\noindent
where $\sigma$ is the effective Hertzian diameter and $\theta(r)$ is the Heaviside function. This is determined by comparing the predicted potential $V_H(r)$ with the numerical one, $V(r)$ (Eq.~\ref{eq:pot}), using the elastic moduli calculated as described in the text. We also compare the numerical data and the Hertzian expression of the microgel-microgel effective interaction with the Tatara theory, which was developed to describe the deformation of a homogeneous elastomeric sphere subject to a large strain~\cite{tatara1991compression}. The theory does not provide a closed expression for the effective force as a function of the strain. However, the latter quantity can be derived by solving the following system of non-linear equations for each separation $r$:~\cite{liu1998large,yan2009mechanical}

\begin{equation}
\label{eq:tatara}
\begin{cases}
\frac{3 (1 - \nu^2) F_T(r)}{2 Y a} - \frac{2 F_T(r) f(a)}{\pi Y} - r + \sigma = 0\\
\left(\frac{3 (1 - \nu^2) \sigma F}{8 Y}\right)^{\frac{1}{3}} - a = 0
\end{cases}
\end{equation}

\noindent
where the two unknowns are the deformation force $F_T(r)$ and the contact radius $a$. The utility function $f(a)$ is given by

$$
f(a) = \frac{(1 + \nu) \sigma^2}{2 (a^2 + \sigma^2)^{3/2}} + \frac{1 - \nu^2}{\sqrt{a^2 + \sigma^2}}.
$$

\noindent
The system of equations~\eqref{eq:tatara} reduces to Eq.~\eqref{eq:hertzian_fit} in the Hertzian limit. Interestingly, a semi-empirical version of the Tatara theory has been used to develop a model that predicts the nonlinear rheology of glasses made of polyelectrolyte microgels~\cite{seth2011micromechanical}.

In order to perform a comparison between the numerical and theoretical data, it is also necessary to consider an additional constant $A$ in the effective potential, that accounts for the (weak) repulsion due to loose dangling ends that stick out of the corona~\cite{boon2017swelling} and whose role will be discussed later on. All in all, we consider the following expression 
\begin{equation}
V(r)=V_H(r)+A.
\end{equation}

During the umbrella sampling simulations we also monitor the shape of the microgel conformations as the separation between the two particles varies. To this aim, we consider the so-called relative shape anisotropy parameter, defined as~\cite{relative_anisotropy} 
\begin{equation}
\label{anisotropy}
\kappa^2 = \left\langle \frac{3}{2}\frac{a_1^2 + a_2^2 + a_3^2}{(a_1 + a_2 + a_3)^2} - \frac{1}{2} \right\rangle
\end{equation}
\noindent
where $\langle \cdot \rangle$ indicates an average over all conformations. The value of $\kappa^2$ is linked to the mass distribution of the monomers: if $\kappa^2 = 0$ the spatial distribution of the monomers is spherically symmetric, whereas if $\kappa^2 = 1$ all the monomers lie on a line.

Finally, we also run bulk simulations of particles interacting through the effective interactions derived above, both $V(r)$ and $V_H(r)$, in order to compare the properties of the two types of systems for different nominal packing fractions $\zeta$. We thus perform molecular dynamics simulations of polydisperse systems (with a polydispersity index of $10\%$ in order to suppress crystallisation) composed of $N = 1000$ particles in the $NVT$ ensemble for several values of the nominal packing fraction $\zeta = \frac{\pi\sigma^3}{6} \frac{N}{V}$, where $\sigma$ as above is the Hertzian effective diameter and $V$ is the volume of the simulation box. We set $k_B T = \frac{1}{\beta} = 1$ and monitor the radial distribution function $g(r)$ and the equation of state, $\beta P = \beta P(\zeta)$, where $P$ is the virial pressure.

\begin{figure*}[t!]
\centering
\includegraphics[width=0.9\textwidth,clip]{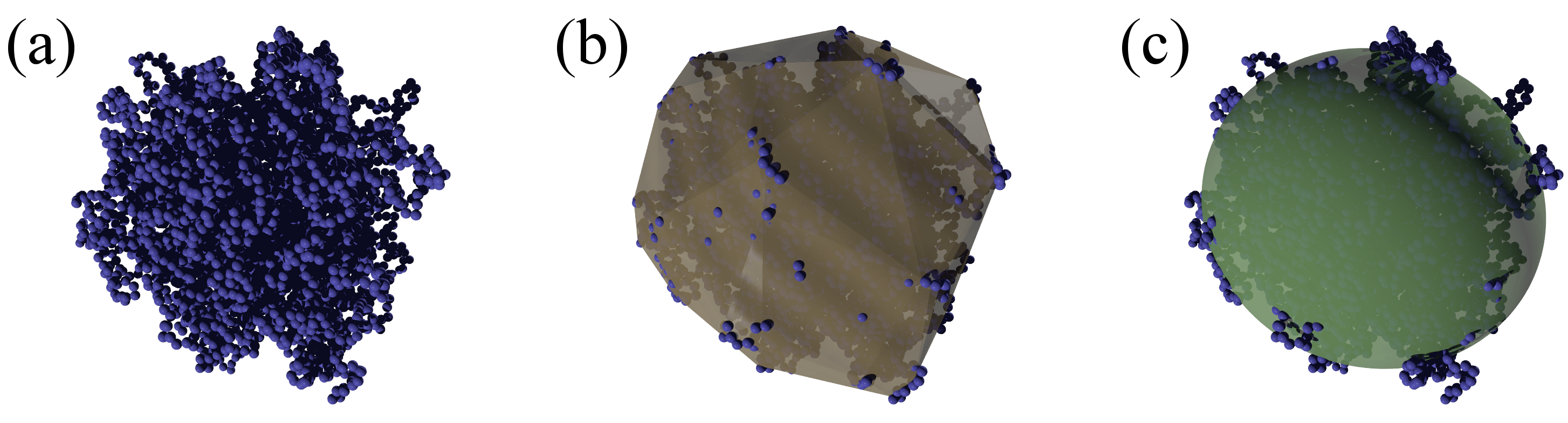}
\caption{\label{fig:cartoon}(a) An equilibrated microgel configuration, (b) its convex hull and (c) its associated ellipsoid.}
\end{figure*}

\section{Generalized method to calculate all single-particle elastic moduli}
\label{sec:elasticity}
In order to probe the elasticity of single particles we build on the method reported in Ref.~\cite{aggarwal_pre} and extend it to the case of disordered polymer networks. The basic idea is to monitor the equilibrium fluctuations of  shape and volume of single microgels and to link their distributions to the elastic moduli. An operative definition for the shape of an instantaneous microgel conformation is therefore required. Since, at the lowest order, thermal fluctuations transform spheres into ellipsoids~\cite{love2013treatise,PhysRevA.36.4371}, we use the latter to approximate the single configurations. Given the heterogeneity of the microscopic structure of microgels, and hence of their mass distribution, the ellipsoid of gyration does not provide a good estimate of the shape. We have thus devised a procedure that does not explicitly depend on the density (or, equivalently, mass) profile of the particle, but only on the positions of the monomers that compose the outer portion of the particle. This is done by applying the following scheme to each equilibrated configuration:

\begin{enumerate}
\item Generate the smallest convex set of points that enclose the microgel configuration, \textit{i.e.} its convex hull.
\item Calculate the gyration tensor of this new set of points using the centres of mass of each triangle composing the convex hull.
\item Diagonalise the gyration tensor, obtaining the three eigenvalues $\lambda_1$, $\lambda_2$ and $\lambda_3$. These are sorted so that $\lambda_1 \geq \lambda_2 \geq \lambda_3$.
\end{enumerate}

The three semi-axes $\lbrace a \rbrace$ are then related to the eigenvalues through the relation $a_i = \sqrt{3\lambda_i}$. Figure~\ref{fig:cartoon} provides a pictorial example of a microgel with crosslinker concentration $c = 5.0\%$, showing the bead-spring configuration, its convex hull and the associated ellipsoid.

The fluctuations in shape of a given microgel configuration are evaluated by constructing the Green-Lagrange strain tensor, $\mathbf{C} = \mathbf{F}^T \cdot \mathbf{F}$, which provides a measure of the local deformation, starting from the deformation gradient tensor $\mathbf{F}$. The latter is defined with respect to a stress-free configuration~\cite{doghri2013mechanics,aggarwal_pre}. Here we choose as reference configuration an ellipsoid of semi-axes $\langle a_1 \rangle$, $\langle a_2 \rangle$, $\langle a_3 \rangle$, where the angular brackets indicate ensemble averages. We thus write the deformation gradient tensor as:
\begin{equation}
\label{eq:deformation_tensor}
\mathbf{F} = \begin{pmatrix}
\frac{a_1}{\langle a_1 \rangle} & 0 & 0\\
0 & \frac{a_2}{\langle a_2 \rangle} & 0\\
0 & 0 &\frac{a_3}{\langle a_3 \rangle}\\
\end{pmatrix}
\end{equation}
\noindent which implies

\begin{equation}
\mathbf{C} = \begin{pmatrix}
\left( \frac{a_1}{\langle a_1 \rangle} \right)^2 & 0 & 0\\
0 & \left( \frac{a_2}{\langle a_2 \rangle} \right)^2 & 0\\
0 & 0 & \left( \frac{a_3}{\langle a_3 \rangle} \right)^2\\
\end{pmatrix}
\end{equation}
We thus build the following three strain invariants~\cite{doghri2013mechanics}:
\begin{align}
J & = \sqrt{\det(\mathbf{C})}\\
I_1 & = \mathrm{tr}(\mathbf{C}) J^{-2/3}\\
I_2 & = \frac{1}{2} \left[ \mathrm{tr}^2(\mathbf{C}) - \mathrm{tr}(\mathbf{C}^2) \right] J^{-4/3}.
\end{align}
\noindent
Here $J$ is linked to the volume change, whereas $I_1$ and $I_2$ are connected to the variation of the shape at fixed volume. We analyse each configuration and calculate the probability distribution functions $P(J)$, $P(I_1)$ and $P(I_2)$. By construction, the reference configuration has $\mathbf{C} = \mathbf{I}$, $J_{\rm ref} = 1$ and $I_{{\rm ref}, 1} = I_{{\rm ref}, 2} = 3$. In addition, since the reference configuration is stress-free, the probability $P(X)$ should be maximum at $X = X_{\rm ref}$. We find that for all investigated cases, the probability distributions always exhibit a maximum at or very close to $X_{\rm ref}$, as shown below, confirming the suitableness of our choice of a reference configuration.

\begin{figure*}[h!]
\centering
\includegraphics[width=0.45\textwidth,clip]{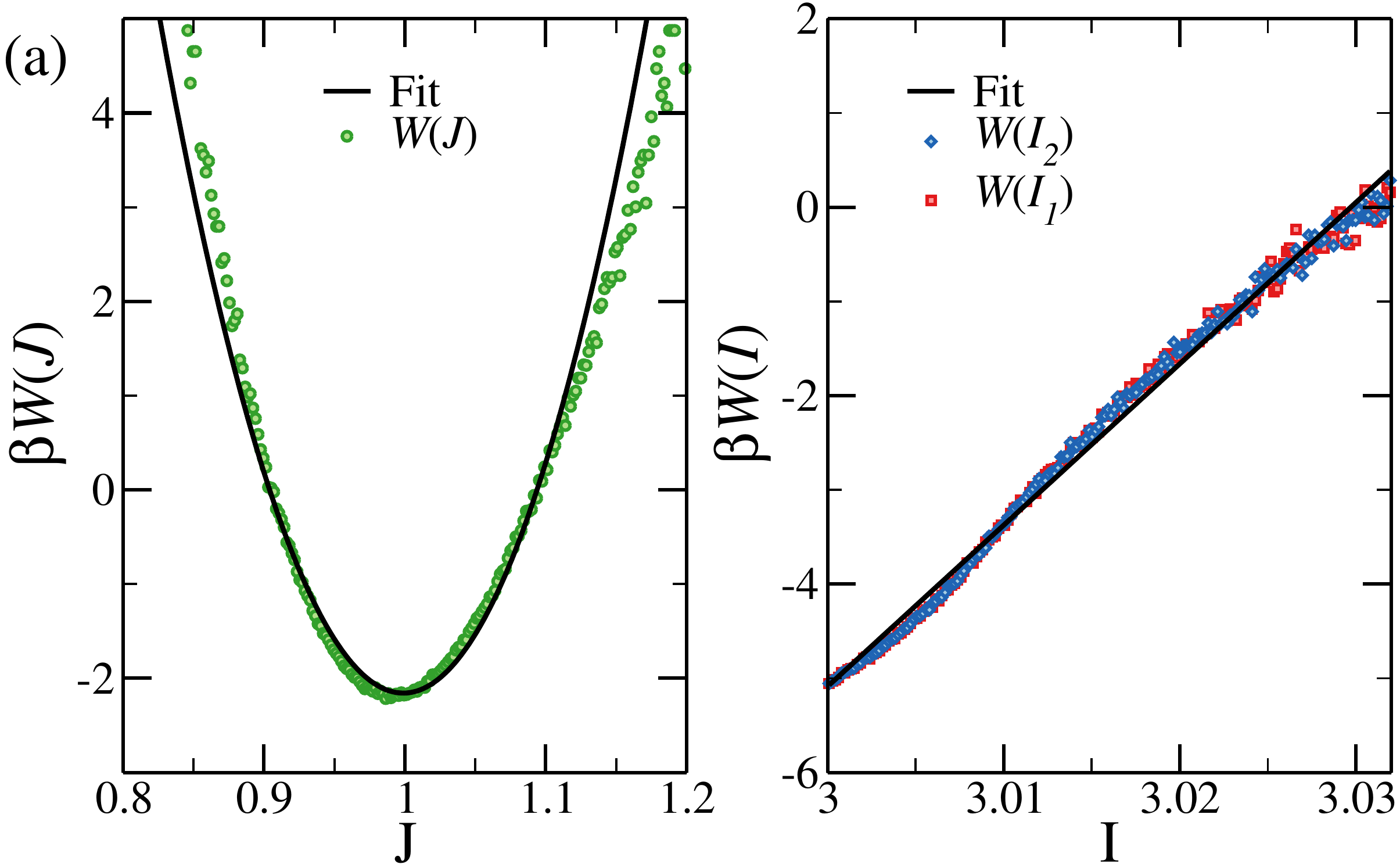}
\includegraphics[width=0.46\textwidth,clip]{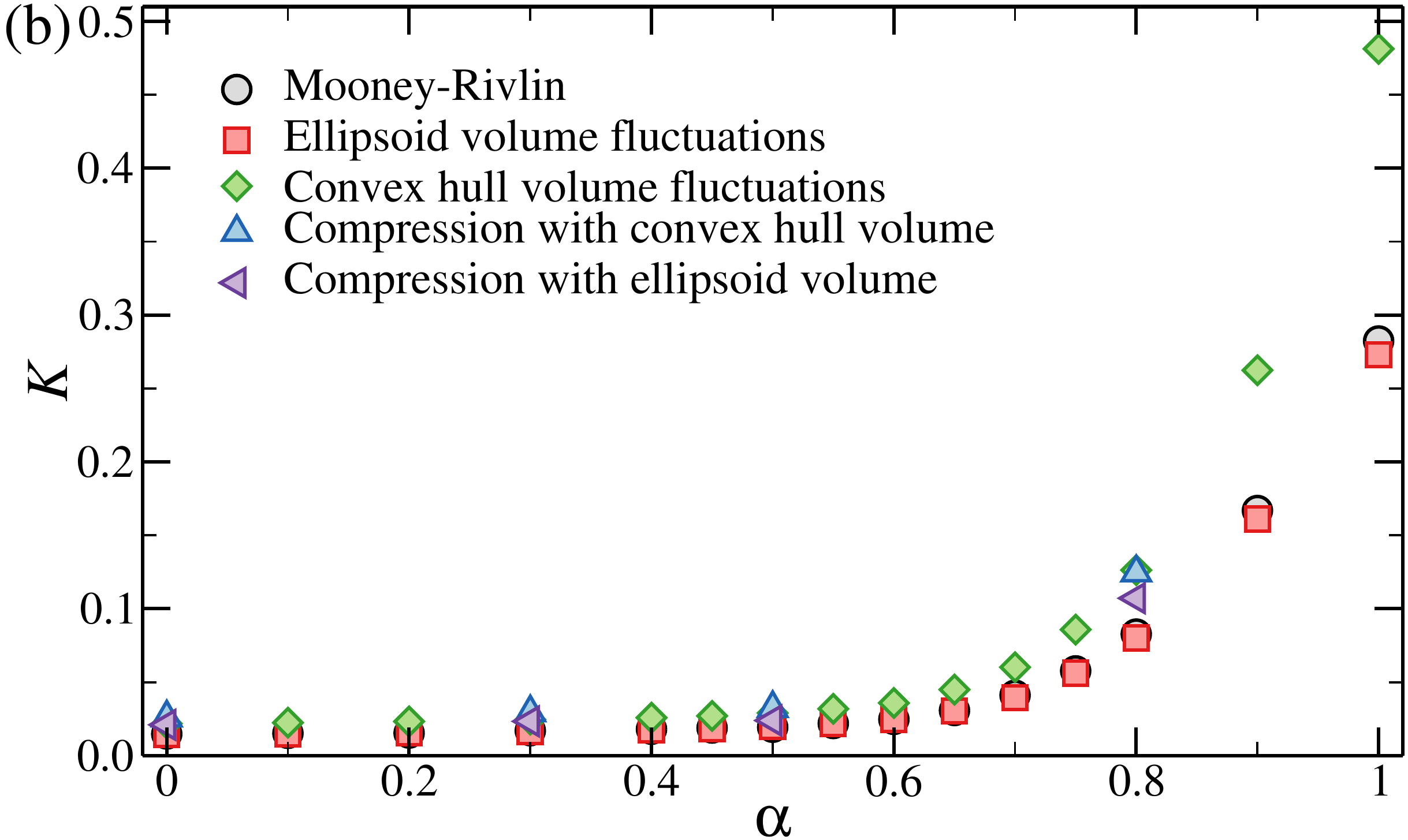}
\caption{\label{fig:pmf_example} (a) Potentials of mean force for (left) $J$ and (right) $I_1$ and $I_2$. Points are simulation data, lines are fits (see text for details); (b) bulk modulus $K$ computed with different methods as a function of $\alpha$.}
\end{figure*} 

By following the phenomenological Mooney-Rivlin theory on rubber elasticity, the energy due to thermal excitations can be written as a function of the strain invariants as~\cite{rubber_elasticity}:

\begin{equation}
\label{eq:energy}
U(J, I_1, I_2)  = U_0 + W(J) + W(I_1) + W(I_2) = U_0 + V \left( \frac{K}{2}(J - 1)^2 + C_{10}(I_1 - 3) + C_{01}(I_2 - 3) \right)
\end{equation}
\noindent
where $U_0$ is the energy of the reference configuration, $V = \frac{4}{3} \pi \langle a_1 \rangle \langle a_2 \rangle \langle a_3 \rangle$ is the microgel volume, $K$ is the bulk modulus and $2(C_{10} + C_{01}) = G$, with $G$ being the shear modulus. In Ref.~\cite{aggarwal_pre} it was assumed that, since the deformations are statistically independent,the three $W$ functions can be approximated with the potentials of mean force (PMFs) extracted from the respective probability distribution functions, \textit{viz.}

\begin{equation}
W(X) = -k_B T \ln P(X) + D_X
\end{equation}
\noindent
with $X = J$, $I_1$ or $I_2$ and $D_X$ an arbitrary constant. Here we first check the assumption that the three strain invariants are statistically independent. We do so by computing the Spearman correlation coefficient $\rho_s$~\cite{norman2008biostatistics} between the time series extracted from the simulations. Regardless of the microgel and of the value of $\alpha$, we always find $\rho_s(I_1, I_2) > 0.999$ and $|\rho_s(I_i, J)| < 0.1$ ($i = 1, 2$). These numbers show that, for all the systems considered, shape and volume fluctuations are always decorrelated. By contrast, the two types of shape fluctuations, embodied by $I_1$ and $I_2$ are always perfectly correlated. We thus find that, under equilibrium conditions, $I_1 \approx I_2 = I$ and hence Eq.~\ref{eq:energy} becomes

\begin{equation}
\label{eq:energy_new}
U(J, I)  = U_0 + W(J) + W(I) = U_0 + V \left( \frac{K}{2}(J - 1)^2 + (C_{10} + C_{01})(I - 3)\right).
\end{equation}
\noindent
It follows that from the analysis of the shape and volume fluctuations it is not possible to obtain the individual values of $C_{10}$ and $C_{01}$, but only their sum.

Fitting the potentials of mean force to functions of the form $M_X (X - X_0)^\gamma + D_X$, where $\gamma = 2$ for $X = J$ and $\gamma = 1$ for $X = I$, yields the elastic moduli through the relations:

\begin{align}
K & = \frac{2 M_J}{V}\\
G & = \frac{2 M_I}{V}.
\end{align}
All other elastic moduli (and related quantities) can be expressed as functions of $K$ and $G$~\cite{doghri2013mechanics}. For instance, the Young's modulus $Y$ and the Poisson's ratio $\nu$ are given by
\begin{align}
\label{eq:Y_def}
Y & = \frac{9 K G}{3K + G}\\
\label{eq:nu_def}
\nu & %= \frac{3 K - Y}{6 K} 
= \frac{3K - 2G}{2(3K + G)}.
\end{align}

Figure~\ref{fig:pmf_example}(a) shows an example of PMFs and related fits for a microgel with $c = 5.0\%$. The volume PMF $W(J)$ is always centered very close to one, whereas the shape PMFs $W(I_1)$ and $W(I_2)$ are monotonically increasing functions of their argument and overlap perfectly with each other, in agreement with the analysis of the correlation coefficients reported above. All PMFs can be well-fitted by the corresponding theoretical curves in a region of at least $2 k_B T$ above the minimum. Farther away $W(J)$ becomes slightly skewed, whereas the    two shape PMFs develop a shoulder and then keep following a straight line with a slope that, within the numerical noise, is compatible with the initial slope. This behaviour is shared by all microgels investigated in this work.

To validate the approach outlined above, we calculate the bulk modulus $K$ in four other ways and compare it to the one calculated from $W(J)$. Two of these alternative methods exploit the fact that $K$ is the inverse of the isothermal compressibility $\chi_T$, which can be straightforwardly computed from the volume fluctuations as~\cite{hansen2013theory}:
\begin{equation}
\label{eq:compressibility}
\chi_T = k_B T \frac{\langle V^2 \rangle - \langle V \rangle^2}{\langle V \rangle}
\end{equation}
\noindent
where $V$ is the instantaneous volume. We estimate $V$ either as the volume enclosed by the convex hull or as the volume of the ellipsoid that best approximates it.

The bulk modulus can also be evaluated by directly looking at the elastic response of the microgels upon compression. To do so, we confine microgels in spherical cavities of decreasing radius and estimate the resulting internal pressure $P_{\rm int}$. 
For small compressions $P_{\rm int}$  linearly depends on $V$, with $K$ being the proportionality constant~\cite{nikolov2018mesoscale}. We thus fit $P_{\rm int}(V)$ in the linear regime, where $V$ can be taken as either the convex hull or the ellipsoid volume, to yield two more alternative estimates of $K$.

Figure~\ref{fig:pmf_example}(b) compares the different results for $K$ for a microgel with $c=5\%$ as a function of the solvophobicity parameter $\alpha$, which plays the role of the temperature, as obtained by each of the five methods just described. First of all, we notice that data sets closely follow each other as $\alpha$ varies. It is evident that the Mooney-Rivlin approach and the ellipsoid-volume fluctuations yield essentially the same results, since both are based on the same assumptions and make use of the same input data. By contrast, the fluctuations of the convex-hull volume and the %$P_{\rm int}$ vs $V_{\rm ext}$ 
compression approach result in somewhat larger values of $K$. This difference is due to the fact that approximating the convex hull with an ellipsoid results in slightly larger fluctuations of the microgel volume and hence in a smaller $K$. Depending on the considered microgel, the observed difference in $K$ between the various methods ranges from $\approx 20\%$ to $\approx 50\%$. In particular, the difference between the Mooney-Rivlin approach and the %$P_{\rm int}$ vs $V_{\rm ext}$ 
compression method is even smaller when the volume of the ellipsoid is considered. Given that there is no unambiguous way of determining the characteristic volume of a polymeric object, as different applications may need different definitions~\cite{hadizadeh_volume}, it is hard to tell \textit{a priori} which method is best. However, the various sets of curves can be readily rescaled on top of each other by means of a multiplicative shift, and they all lead to qualitatively similar results for all investigated cases. For the above reasons, in the following we focus on the Mooney-Rivlin approach, because this is the only method that allows us to simultaneously obtain also the shear modulus $G$, and from these two moduli, to extract all the others.

\section{Results}
\subsection*{Single-particle elastic moduli of neutral microgels}

\begin{figure*}[h!]
\includegraphics[width=0.41\textwidth,clip]{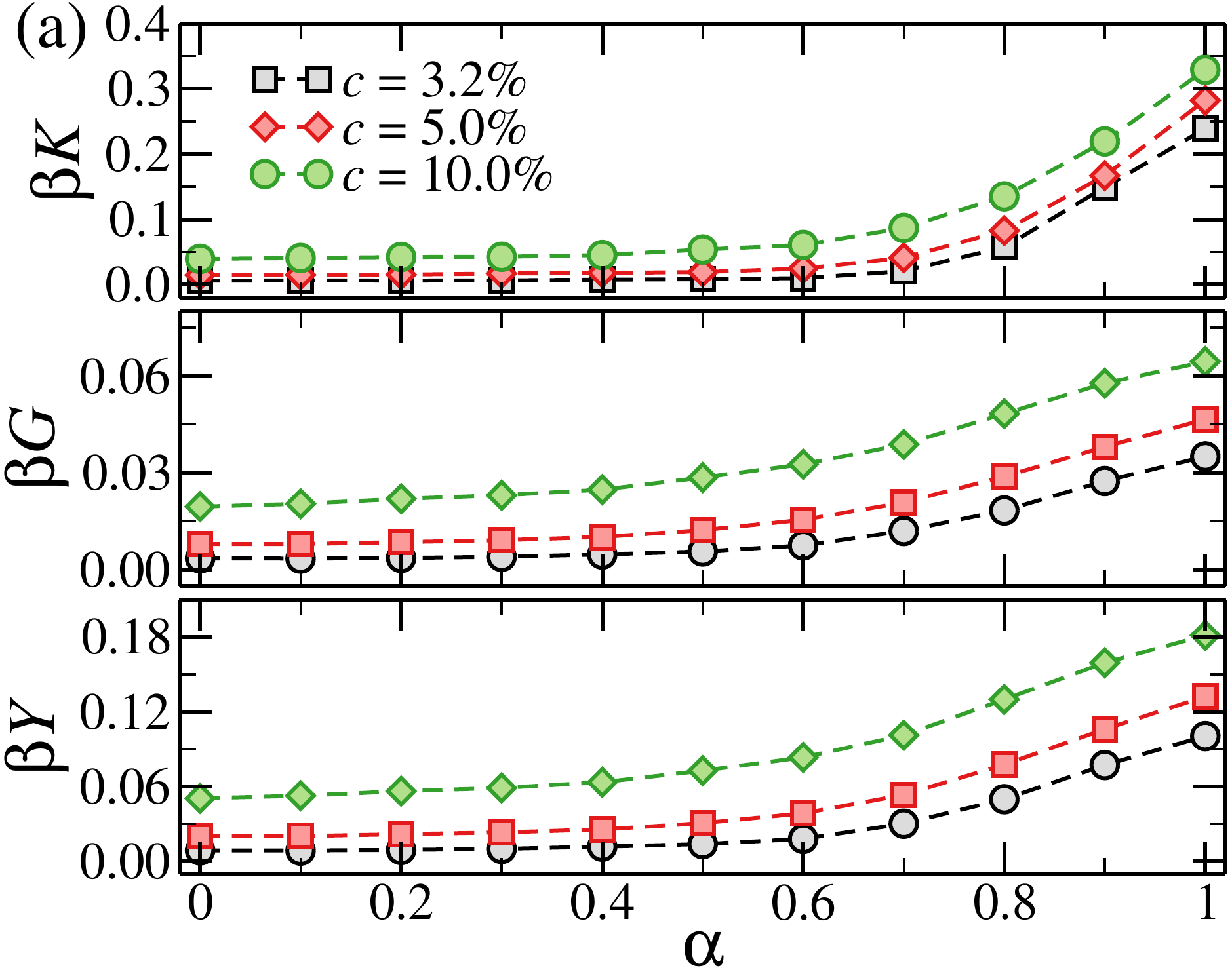}
\hspace{0.2cm}
\includegraphics[width=0.53\textwidth,clip]{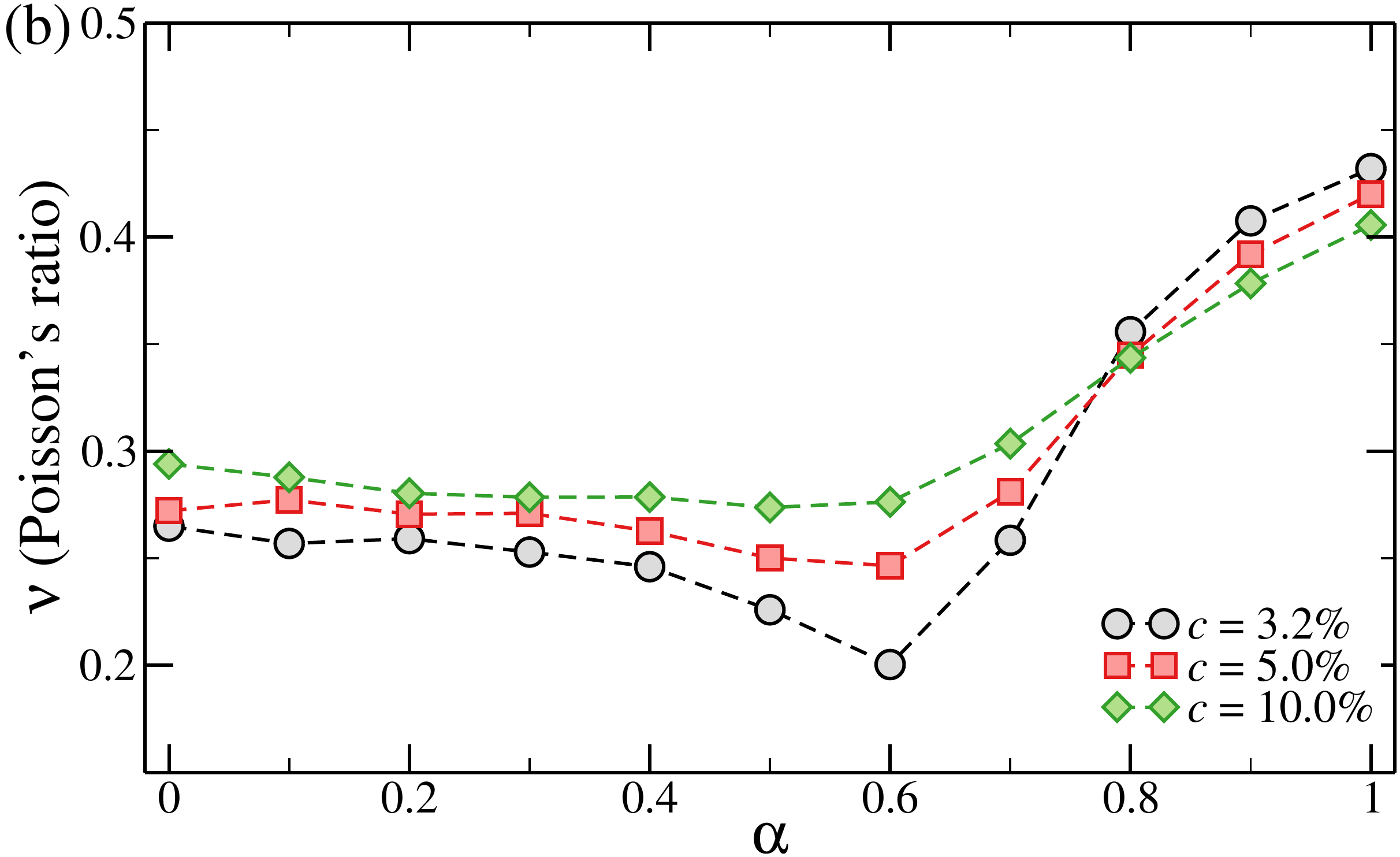}

\caption{\label{fig:moduli} (a) The elastic moduli (top) $K$, (middle) $G$ and (bottom) $Y$ (normalised by $\beta = 1/k_BT$) as functions of $\alpha$ for all investigated microgels. The elastic moduli are expressed in units of $1 / \sigma_m^3$; (b) Poisson's ratio $\nu$ for the three microgels as a function of $\alpha$. }
\end{figure*}

Having established and validated a robust procedure to calculate the moduli of single microgels, we now apply this methodology to microgels with three different crosslinker concentrations, $c=3.2\%, 5\%$ and $10\%$, for different values of the solvophobicity parameter $\alpha$ across the volume phase transition. 

Figure~\ref{fig:moduli}(a) reports the results for the bulk, shear and Young's moduli for three different microgels as a function of $\alpha$ from the fully swollen ($\alpha = 0$) to a deswollen (albeit not fully collapsed) state ($\alpha = 1.0$). For all cases investigated here all the elastic moduli  increase monotonically with $\alpha$ and $c$, at least within the numerical noise. The increase with $c$ is readily explained by noting that the elastic moduli are monotonically increasing functions of the (effective) crosslinker concentration~\cite{rubinstein2003polymer}. However, we note that, contrary to simple scaling arguments~\cite{de1979scaling}, the bulk and shear moduli we obtain do not depend linearly on the number of chains (and hence of crosslinkers).
The monotonic dependence of the elastic moduli on $\alpha$ is more puzzling. Indeed, experimental results for hydrogels~\cite{hirotsu_hydrogels} and microgels of different size~\cite{hashmi2009mechanical,voudouris2013micromechanics} seem to suggest that the bulk modulus, the Young's modulus or both should exhibit a dip close to the VPT temperature. Such a behaviour might be explained, at least from a qualitative standpoint, by noting that $K$ is the inverse of the isothermal compressibility $\chi_T$. Even though it is still not clear whether the VPT in microgels should be modelled as a proper phase transition or as a crossover~\cite{habicht2014non,habicht2015critical}, $\chi_T$ should always display a maximum at the transition (or crossover) temperature. Nevertheless, we do not observe any minimum in $K$. One possible reason for this behavior could be the small size of the investigated microgels, while another reason could be the fact that we use an implicit solvent. While we intend to investigate these two aspects in the future by examining larger system sizes and microgels in an explicit solvent\cite{camerin2018modelling}, we also note that 
a recent numerical study on relatively large microgels in an explicit solvent reported a bulk modulus that also monotonically increases across the VPT, in qualitative agreement with our results~\cite{nikolov2018mesoscale}.
About the dependence of the Young's modulus on $\alpha$, there is in principle no evident reason why it should display a minimum at the VPT.  

Thanks to the joint calculation of two different elastic moduli, we are able to numerically quantify the Poisson's ratio $\nu$ of \textit{in silico} microgels. Figure~\ref{fig:moduli}(b) shows $\nu$ for the three investigated microgels across the VPT. We first of all note that, for the range of parameters investigated here, $\nu$ is in good agreement with that measured for PNIPAM gels~\cite{hirotsu_hydrogels} and predicted by polymer theories~\cite{boon2017swelling}, whereas in the swollen regime it is smaller than the one measured for large microgels~\cite{voudouris2013micromechanics}. Interestingly, as the microgel deswells a clear non-monotonic behaviour is observed, with a minimum appearing in the range $\alpha \in [0.5, 0.6]$, \textit{i.e.} close to the VPT temperature. We find that the minimum deepens as the crosslinker concentration decreases. Unfortunately, to our knowledge there is no available data on the Poisson's ratio of microgels with a submicrometer size, and thus we have to resort to draw a comparison with materials that have a markedly different inner architecture with respect to our microgels. Despite the qualitative difference observed in the individual behavior of the $K$, $G$ and $Y$ moduli, the appearance of a minimum in the Poisson's ratio is in agreement with the available experimental results on PNIPAM-based materials. Indeed, this feature has been observed both for hydrogels~\cite{hirotsu_hydrogels} and for microgels of diameter $\approx 110$ $\mu$m investigated with capillary micromechanics~\cite{voudouris2013micromechanics}, the latter being usually much more homogeneous than their submicron counterparts.
In the literature, the emergence of a minimum in the Poisson's ratio has always been explicitly or implicitly presumed to be connected with the presence of a minimum in either the bulk or Young's modulus~\cite{hashmi2009mechanical,voudouris2013micromechanics}. By contrast, our results clearly show that the two phenomena are not necessarily linked and that a genuine minimum is found also when all other moduli are monotonic.

It is now interesting to compare the calculated moduli with the available experimental measurements. First of all, in agreement with experiments, we find that $K > G$. Converting our numerical units to the experimental ones by considering microgels with diameters of the order of $\approx 400$~nm, we find that the measured moduli range from tens to hundreds of Pascals (or even thousands as $\alpha$ increases), increasing with crosslinker concentration $c$. Micromechanical measurements~\cite{voudouris2013micromechanics} yield $K$ in the range $10^3-10^4$ Pa and $G$ in the range $10^2-10^3$ Pa in the swollen state for particle of radius from $~20$~$\mu$m to $~50$~$\mu$m. The moduli are also found to increase with $c$. In addition, from microrheology measurements for microgel of 1$\mu$m size, a value of the shear modulus per particle was extrapolated in the range of $10^2$ Pa~\cite{di2013macro} in closer agreement with our numerical estimates. On the other hand, Atomic Force Microscopy (AFM) measurements for microgels with a conventional radius of the order of 300~nm reported $Y$ of the order of $10^4$ Pa\cite{hashmi2009mechanical}. This value is higher than what would be estimated from the micromechanics measurements discussed above by using Eq.~\ref{eq:Y_def}, suggesting that the Young's moduli extracted from AFM may be sensibly larger than the real ones. This may be due to the well-known dependence of $Y$ on the indentation depth\cite{burmistrova2011effect}, meaning that AFM measurements may actually provide the Young's modulus of the core of the microgel, rather than that characteristic of particle, as the method may not be sensitive enough to carefully measure the weak corona contribution.

To summarize the results of this section, we note that we have been able to provide an overall realistic estimate of single particle moduli. However, a systematic investigation by changing particle size and a careful comparison between different measurements will be crucial to further elucidate this point. It is also important to stress that our approach features the common flexible bead-spring model for polymers, and it has not been adjusted to describe in particular PNIPAM or another specific polymer. Employing more refined models (such as semi-flexible polymers or an explicit solvent) will provide other sources of refinement in order to achieve a better quantitative description of experiments. All in all, the minimum that we find in the Poisson's ratio close to the VPT is a remarkable result, being the first numerical confirmation of a general experimental trend found in microgel suspensions as well as in hydrogels, which confirms the robustness and the validity of the methodology developed in this work.

\subsection*{Effective interactions and validation of the Hertzian model at small deformations}

\begin{figure*}
\centering
\includegraphics[width=0.85\textwidth,clip]{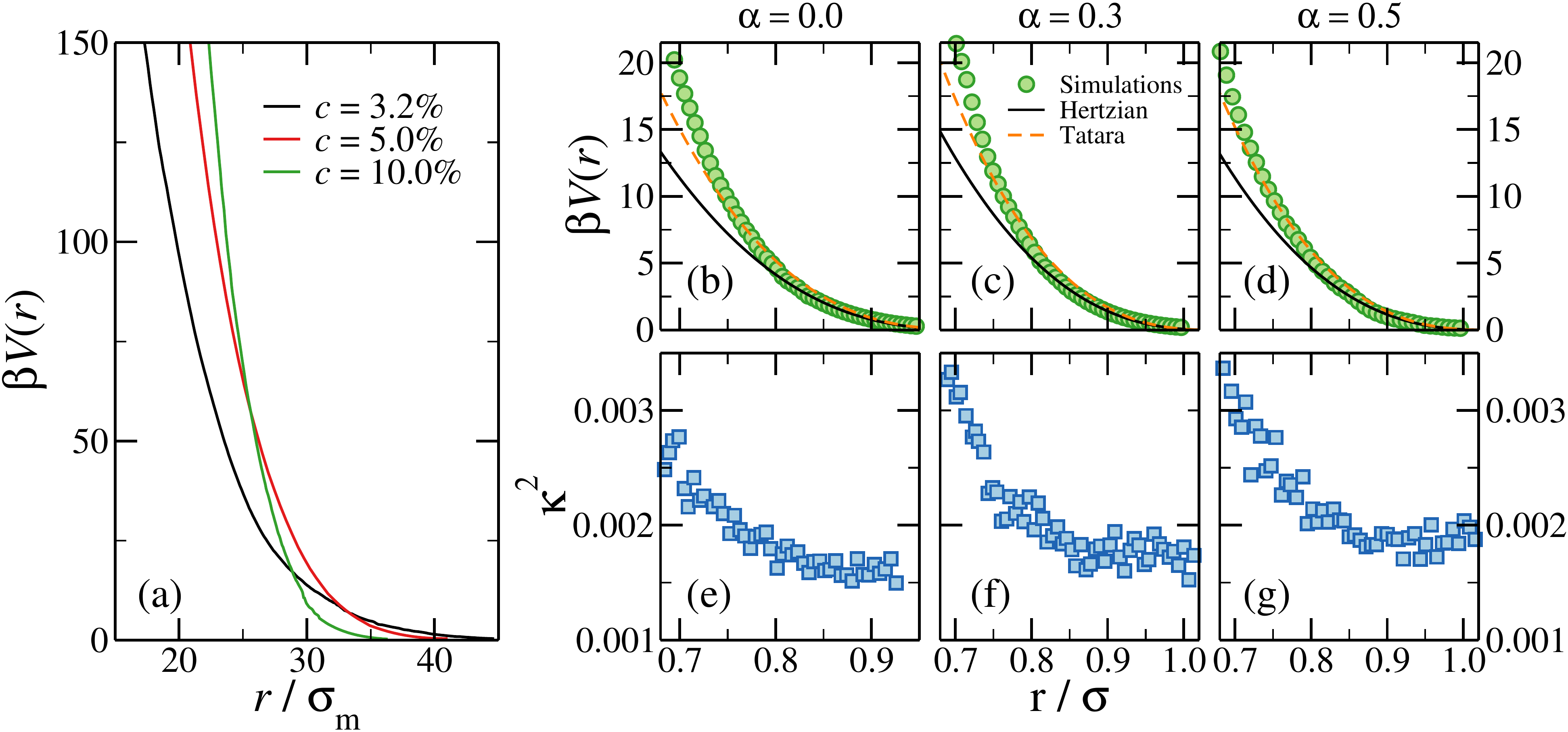}
\caption{\label{fig:fits} (a) Effective interactions $\beta V(r)$ between two microgels with various crosslinker concentrations in the swollen state ($\alpha = 0.0$) shown as a function of the distance between them normalized by the monomer size, $r/\sigma_m$; (b-d) comparison between the calculated $\beta V(r)$ (symbols), the Hertzian potential (solid lines) and the Tatara potential (dashed lines) as a function of the microgel-microgel separation $r/\sigma$ for microgels with $c = 5.0\%$ and three different values of $\alpha$ approaching the VPT from below. The amplitude of the Hertzian potential is fixed by the values of the calculated elastic moduli (see~\eqref{eq:hertzian_fit});
(e-g)  Corresponding shape anisotropy parameter $\kappa^2$ (see~\eqref{anisotropy}).}
\end{figure*}

We now report results of the effective two-body interaction $\beta V(r)$, calculated from explicit simulations of two microgels. Figure~\ref{fig:fits}(a) shows the calculated potentials for the three studied crosslinker concentrations in the swollen state ($\alpha=0.0$). As expected, the potential is always repulsive and becomes steeper for increasing $c$, while its range decreases because microgels become more compact. We compare the calculated interactions with the Hertzian model, whose strength is fixed to the elastic moduli that we have calculated. Thus, the only free parameters involved in the fitting procedure are the effective Hertzian diameter $\sigma$ and the residual repulsion $A$.
Figure~\ref{fig:fits}(b-d) show the comparisons of $\beta V(r)$  for the $c = 5.0\%$ microgel with the Hertzian predictions, \eqref{eq:hertzian_fit}, for three different values of $\alpha$.  Data are shown only up to $20$~$k_BT$ because we find that the Hertzian fits work well but only in the region $0.5 \leq \beta V(r) \leq 6$. Remarkably, we obtain a good agreement with simulation data for all microgels investigated and hence for all examined crosslinker concentrations. We thus confirm the Hertzian nature of the microgel-microgel effective interactions in good solvent, at least for average separations that are associated to an effective repulsion of the order of a few times the thermal energy. The high quality of the fits also demonstrates the validity of the method we use to extract the values of the elastic moduli from the simulations. Indeed, we checked that values of $Y$ that differ by a factor of two or more generate deviations from the numerical effective interactions that cannot be cured by varying $\sigma$ or $A$.

The origin of the deviation from the Hertzian behaviour can be traced back to the breakdown of one of the key assumptions of the CET. When microgels start to overlap more strongly with each other, and hence feel a mutual repulsion that far exceeds the thermal energy, they also deform. This can be seen in panels (e-g) of Figure~\ref{fig:fits}, which shows the relative shape anisotropy parameter $\kappa^2$ as a function of the separation $r$. Regardless of the value of $\alpha$, we see that $\kappa^2$ remains fairly constant down to a value of $r$ below which it starts to increase more rapidly. This crossover value of $r$ is, for all the cases investigated here, fully compatible with the distance at which the Hertzian and the full effective interaction start to deviate, clearly demonstrating the close connection existing between the extent of the deformation and the resulting effective interaction.

Panels (b-d) of Figure~\ref{fig:fits} also contain a comparison with the Tatara effective interaction computed by solving, for each separation $r$, the system of equations~\ref{eq:tatara}. As a general trend, the agreement between theory and simulations improves at small separations, reaching in some cases $10-20$ $k_BT$. Since the Tatara theory does not require any additional parameter with respect to the Hertzian one, our results indicate that using the former in place of the latter may increase the accuracy of numerical simulations of soft objects.
\begin{figure*}[h!]
\includegraphics[width=0.5\textwidth,clip]{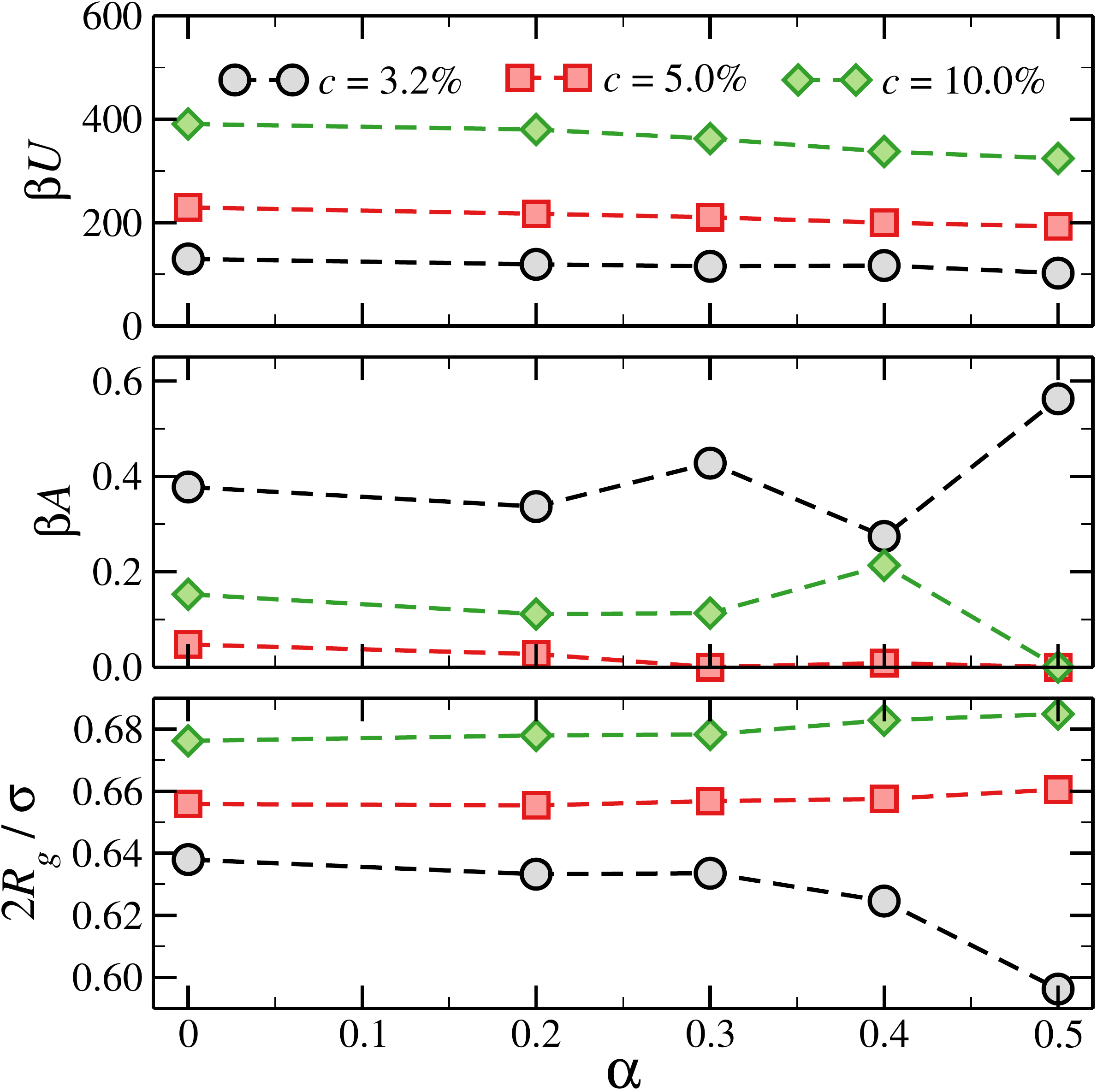}
\caption{\label{fig:U_A_hydro_over_sigma} (Top) The Hertzian energy prefactor, (middle) the dangling-end residual repulsion contribution and (bottom) the ratio between the diameter of gyration $2R_g$ and the elastic diameter $\sigma$ for microgels of varying crosslinker concentration and for different values of $\alpha$.}
\end{figure*}

In addition to the confirmation of the Hertzian nature of the effective interactions at small deformations, the fitting procedure also yields quantities that have a physical significance. In particular, the energetic prefactor in~\eqref{eq:hertzian_fit}, $U \equiv \frac{2 Y \sigma^3}{15 (1 - \nu^2)}$, has been estimated for real PNIPAM microgels under good solvent conditions, for instance by fitting experimental data with simulations~\cite{mohanty2014effective} or by inverting the radial distribution function of a sample in the dilute regime~\cite{zhang2009thermal,yodh_chapter}. In general, the resulting repulsion strengths are found to be of the order of tens to hundreds of $k_BT$'s. The top panel of Figure~\ref{fig:U_A_hydro_over_sigma} shows $U$ for the microgels investigated here as functions of $\alpha$. There is a clear trend with $c$ since, owing to their increased softness, microgels with fewer crosslinkers display smaller values of $U$. We always find values that are of the order of hundreds of $k_BT$'s, in qualitative agreement with the fits to experimental data presented in Ref.~\cite{mohanty2014effective}. We have not investigated microgels with $c < 3.2\%$, and hence cannot directly compare with the results of Refs.~\cite{zhang2009thermal,yodh_chapter}, where $c$ is not explicitly calculated but can be estimated to be $\leq 1\%$. However, a linear extrapolation to this range of crosslinker concentrations yields lower bounds of $U \approx 70$~$k_BT$, in agreement with the experimental values. 

It has been suggested that the core-corona structure of microgels might be complemented by an \textit{outer corona} which comprises those dangling ends and loops that can stick out of the particle~\cite{dulle_dangling_ends,boon2017swelling}. Since the density of this region is very small, its contribution towards the radius of gyration is tiny. However, its effect on the hydrodynamic radius should be relevant. From the standpoint of microgel-microgel interactions, the extent of the repulsion exerted by the loose dangling ends and loops is here embodied by the constant $A$. Figure~\ref{fig:U_A_hydro_over_sigma} shows in the central panel that the magnitude of this repulsion is always smaller than $k_B T$. It remains fairly constant as we approach the VPT, although sometimes for $\alpha = 0.5$ the numerical noise increases enough to make it hard to resolve such a tiny contribution with the required accuracy, resulting in cases where $A \approx 0$. Indeed, we notice that we cannot evaluate effective interactions for $\alpha > 0.5$ because the microgels become too attractive and our numerical methods become increasingly inefficient.
However, in the investigated range of $\alpha$ values, our results strongly suggest that the role played by the dangling ends in determining microgel-microgel repulsions is negligible. We note on passing that the pair potentials computed here do not show any effect due to chain entanglements, which are believed to be important for small microgels with low crosslinker contents in the paste regime~\cite{kunz2018polymer}.

The last parameter that comes out of the fitting procedure described above is the effective (elastic) diameter $\sigma$. We find that $\sigma$ is always significantly larger than $R_g$, thus resembling the hydrodynamic radius $R_H$ of the particles. Experimental studies have reported that in general one has $\frac{R_g}{R_H}\approx 0.6$~\cite{senff1999temperature,senff2000influence}, in good agreement with the present findings. This result validates previous studies, which have shown that the Hertzian model works in describing experimental $g(r)$ if $\sigma$ coincides with the hydrodynamic radius~\cite{mohanty2014effective,maxime_nat_comm}. It would be interesting to check this in more detail in the future by performing a quantitative comparison with experiments. In addition, we notice  that $\sigma$ and the gyration radius $R_g$ are always essentially proportional in the whole $\alpha$-range, as highlighted by the bottom panel of Figure~\ref{fig:U_A_hydro_over_sigma}, which shows $2R_g / \sigma$. Interestingly, this ratio changes by less than $10\%$ going from the densest to the softest microgel. These results provide a way of experimentally estimating $\sigma$ without having to directly measure the microgel-microgel effective interactions. Together with the notion that $A$ is negligible and $U$ can be estimated by measuring the single-microgel elastic moduli, our results suggest that it is possible to derive microgel-microgel interactions in the small-deformation regime from measurements of simple single-particle properties.

\subsection*{Validity of the Hertzian model in bulk suspensions}

\begin{figure*}[h!]
\includegraphics[width=0.6\textwidth,clip]{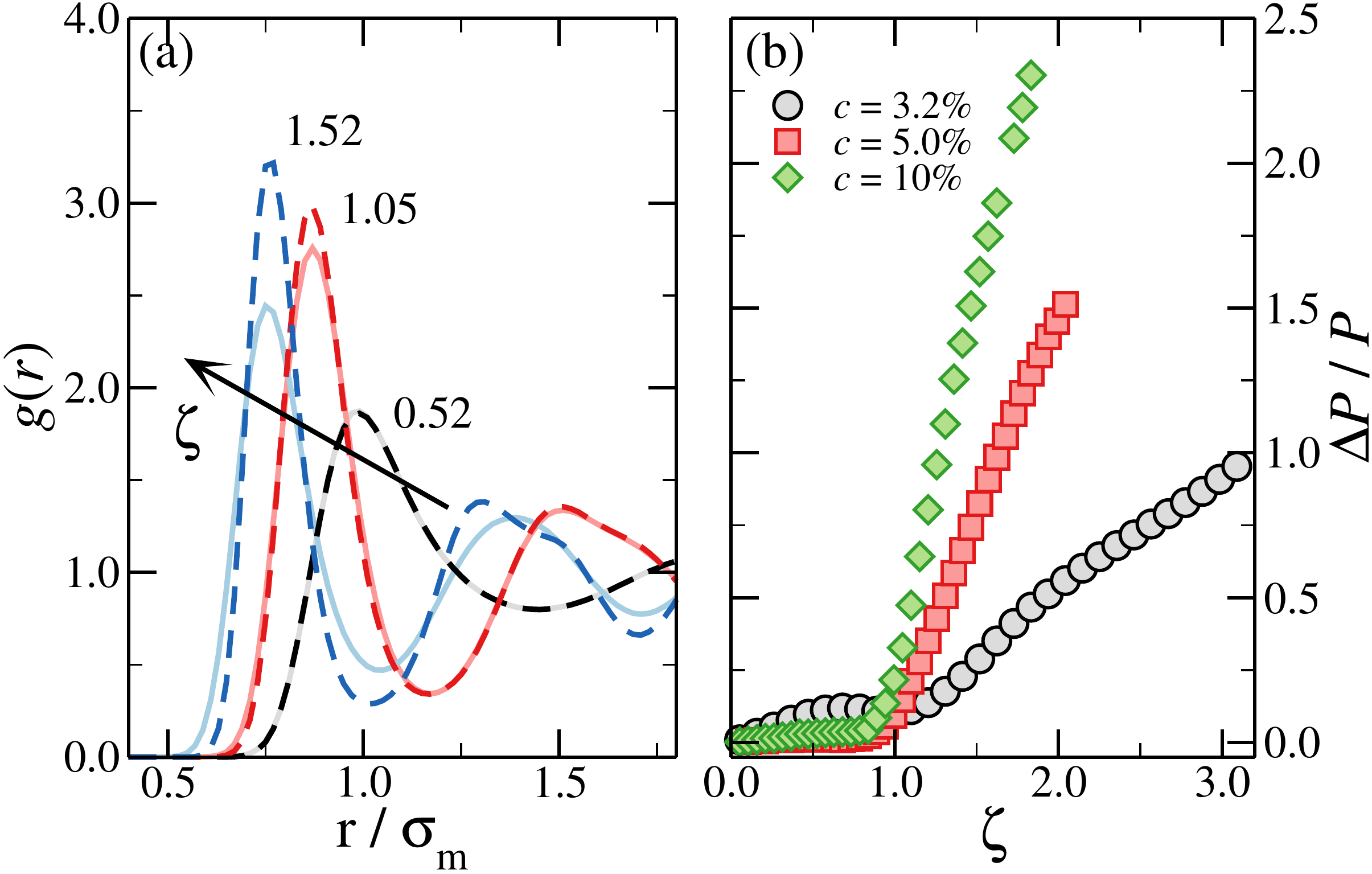}
\caption{\label{fig:gr_P_diff} (a) Radial distribution functions $g(r)$ for particles interacting through the real effective potential (dark dashed lines) and its Hertzian approximation (light solid lines) with $\zeta = 0.52$ (black lines), $\zeta = 1.05$ (red lines) and $\zeta = 1.52$ (blue lines). The system simulated is composed of microgels with $c=5\%$ and $Z=25$~$\sigma_m$. (b) The pressure difference between two systems (normalised by the non-Hertzian pressure) as a function of the packing fraction $\zeta$ for three different microgels.}
\end{figure*}

In the previous section we showed that, outside of the small-deformation regime, where microgels begin to interact more strongly, the Hertzian potential quickly starts to underestimate the repulsion, which gets steeper and steeper as $r$ decreases. At the same time, the shape of the particles becomes less and less spherical, as highlighted by the increase in $\kappa^2$ (see panels (e-g) in Figure~\ref{fig:fits}). As the concentration increases and microgels pack more tightly, many-body interactions acquire a greater importance and a description based on pair potentials becomes questionable~\cite{riest2015elasticity,C6SM01567K}. Thus there exists a crossover in packing fraction $\zeta$ across which the system goes from a Hertzian-dominated regime to a regime where large deformations and the many-body nature of the microgel-microgel interactions come into play.

In order to estimate the packing fraction $\zeta$ at which the Hertzian description breaks down, we simulate bulk systems of microgels modelled as soft spheres interacting through either the calculated effective potentials $V(r)$ of Fig.~\ref{fig:fits}(a) or their Hertzian approximations $V_H(r)$. Figure~\ref{fig:gr_P_diff}(a) shows the radial distribution functions of a system of $c=5\%$ microgels for different $\zeta$. The two potentials we use yield systems with identical structure up to $\zeta \approx 1$. Above this value, the real and Hertzian radial distribution functions start to deviate, with the Hertzian system exhibiting a broadening of the first peak and an overall weakening of the structure compared to the real system. We quantify the dissimilarity between the two systems by evaluating the difference between their equations of state, \textit{viz.}
\begin{equation}
\label{eq:chi_square}
\frac{\Delta P}{P} = \frac{P - P_H}{P}
\end{equation}
\noindent
where $P_H$ and $P$ are the pressures in the Hertzian and non-Hertzian system, respectively. Fig.~\ref{fig:gr_P_diff}(b) shows the dependence of $\Delta P / P$ on the packing fraction for microgels with three different values of $c$. The behaviour of the pressure difference convincingly confirms the structural analysis discussed above: as soon as $\zeta \approx 1$, $\Delta P / P$ becomes much more sensitive to the packing fraction, growing steeply with $\zeta$. At the smallest crosslinker concentration investigated here ($c = 3.2\%$) we observe a deviation at small packing fractions that is most likely ascribable to the effect of the dangling ends which, as seen in Figure~\ref{fig:U_A_hydro_over_sigma}, is more prominent at this value of $c$.

The crossover value at which non-Hertzian contributions start to play a role seems to depend systematically on $c$, with more crosslinked microgels showing an earlier deviation from the Hertzian model. Indeed, we find that the Hertzian model seems to work quantitatively up to $\zeta \approx 0.8$ for $c=10\%$ and up to $\zeta \approx 1.2$ for $c=3.2\%$, indicating that softer microgels obey the Hertzian model in a more extended regime with respect to hard ones. However, we stress that these specific values may depend on the inner architecture and chemical composition of the microgels under study. For the case considered here, the onset of deviations from the Hertzian model takes place for packing fractions at which a two-body effective interactions becomes questionable. Under this regime, jamming transitions~\cite{zhang2009thermal} and density-induced deswelling~\cite{deswelling_density}, which may further alter the inter-microgel interaction, are observed in some microgel systems, underlining the importance of using detailed (many-body) mesoscopic models to investigate high-density conditions.
\section{Discussion and Conclusions}

Polymer-based particles such as star polymers, dendrimers and microgels have become prominent in the soft matter field as a result of the tunability of their internal structure and, consequently, of the control achievable on their mutual interactions. Such a high degree of control comes at the cost of an increase in the complexity of the single-particle properties (\textit{e.g.} size, shape, elasticity), which become more strongly dependent on the external parameters such as temperature or packing fraction. It is thus imperative to establish a link between the microscopic architecture, the particle properties at the mesoscopic level and the macroscopic bulk behaviour. A first attempt to provide such connection by reducing the complexity of the system comes from the Hertzian theory, which can be used to describe soft colloids as elastic spheres whose interaction is determined solely by the single-particle elastic moduli. Despite the large use of such approximation to describe the collective behaviour of soft colloids, a systematic study to establish its validity and failure across several concentrations had not been carried out yet. Here we fill this gap thanks to the joint calculation of single-particle elastic moduli and effective interactions, providing evidence of the validity of the Hertzian model at small deformations and in an extended range of packing fractions in bulk suspensions of soft colloids.

To achieve this goal we exploit a mesoscopic, realistic description of microgel particles, which are being increasingly used as model systems in soft matter, developing a methodology that makes it possible to numerically evaluate the single-particle elastic moduli from the fluctuations of the shape and volume of the particles. This advanced technique is generic for soft particles with complex architecture and can thus be readily extended to other types of systems, such as star polymers~\cite{grest1996star} or dendrimers~\cite{lee2005designing}.

When applying our method to the calculation of the elasticity of single non-ionic microgel particles, we find results in qualitative agreement with the scarce experimental literature. In particular, we observe a non-monotonic behaviour of the Poisson's ratio, with a minimum occurring close to the VPT, as also found in experiments~\cite{voudouris2013micromechanics}. By monitoring the dependence of the elastic properties on the crosslinker concentration, we also show that the minimum becomes deeper with decreasing $c$, a feature that will be interesting to compare with experimental results when more data will be available. Furthermore, we find that our \textit{in silico} microgels display an enhanced stiffness with increasing temperature, originated by the monotonic behavior of the bulk, Young's and shear moduli.

We then deploy different computational techniques, namely umbrella sampling and the generalised Widom insertion scheme, to calculate effective interactions occurring between two microgel particles in a wide range of separation distances. We use the values of the elastic moduli, extracted beforehand, to verify the validity of the Hertzian model from microscopic principles. Indeed, by comparing the real pair interaction with the theoretical predicted one, we find an excellent agreement at large particle distances, up to a repulsion energy of $\approx 6 k_BT$, where microgels pay a small energy penalty to undergo tiny deformations. For shorter separations, the deformation of the microgels becomes important and a clear deviation of the calculated interactions from the Hertzian model is detected, with the former exhibiting a much steeper increase as the distance decreases. This is in agreement with the CET assumptions underlying the Hertzian model, which should indeed be valid only in the small deformation regime. 
To further quantify the validity of the Hertzian pair interactions to describe the behavior of bulk suspensions, we additionally perform simulations of particles interacting through the Hertzian and the numerical pair potential, demonstrating that neutral microgels behave like Hertzian spheres up to nominal packing fractions of the order of $\zeta \approx 1$. Even though more refined theories that also take into account large deformations, such as the Tatara theory~\cite{tatara1991compression}, may extend the range of validity of this type of simulations, in the concentrated regime a pair-potential description becomes debatable~\cite{riest2015elasticity}. In addition, other phenomena that cannot be described by the classical elasticity theory, such as faceting, interpenetration and inter-particle entanglement\cite{mohanty2017interpenetration,Gaurasundar2017} become more and more relevant.
We also notice that the Hertzian approximation could break down also far from too dense conditions, when the interactions are indirectly probed at small particle-particle distance, as in the case of the addition of a depletion agent. Indeed, a recent work has experimentally tested depletion interactions in microgel mixtures, showing that there is the need to go beyond the simple Hertzian model when adding an effective attraction on top of the microgel-microgel interactions, even at relatively low volume fractions\cite{maxime_nat_comm}.

Our work have thus convincingly demonstrated that there exists an intimate link between the elastic properties and bulk behavior of soft colloids only in the regime of small deformations, where the Hertzian model is found to work. While this seems an obvious result, there is a large amount of literature extending the linear elasticity assumptions up to very dense conditions by relying on simple Hertzian or equivalent models. Here we establish the packing conditions beyond which this approach is doomed to fail. In this regime, the microscopic details of the inner structure of the particles become extremely important and must be taken into account in order to achieve a quantitative description of the bulk behaviour of suspensions of soft colloids. %This finding entails that many results on the glass and jamming transition obtained at high density with simple Hertzian or similar soft repulsions should be reconsidered under a new light. 
In the future, more refined approaches, either built on a coarse-grained treatment including many-body interactions\cite{li2018role,doukas2018structure} or that explicitly include the deformability of the particles\cite{gnan2018microscopic,urich2016swelling} will need to be employed in order to correctly describe the peculiar features of soft colloids close to or above close packing~\cite{rovigatti2019numerical}.

\section*{Acknowledgements}

We acknowledge support from the European Research Council (ERC Consolidator Grant 681597, MIMIC). 

\bibliography{library}

\end{document}